\input{aipcheck}

\documentclass[
    ,final            
  ]
  {aipproc}

\layoutstyle{6x9}

\newcommand{\be}{\begin{equation}}
\newcommand{\ee}{\end{equation}}
\newcommand{\bea}{\begin{eqnarray}}
\newcommand{\eea}{\end{eqnarray}}

\newcommand{\bk}{{\bf k}}

\newcommand{\bq}{{\bf q}}

\begin{document}

\title{Exclusive central production of dijets at hadronic colliders}


\author{I.P.~Ivanov}{
  address={IFPA, Universit\'e de Li\`ege, Belgium}
  ,altaddress={Sobolev Institute of Mathematics, Novosibirsk, Russia}
}

\author{J.-R.~Cudell}{
  address={IFPA, Universit\'e de Li\`ege, Belgium}
}

\author{A.~Dechambre}{
  address={IFPA, Universit\'e de Li\`ege, Belgium}
}

\author{O.~Hernandez}{
  address={McGill University, Montreal, Canada}
}

\begin{abstract}
We critically re-examine the calculation
of central production of dijets in quasi-elastic hadronic
collisions. We find that the process is not dominated by the
perturbative contribution,
and discuss several sources of uncertainties in the calculation.
\end{abstract}

\maketitle


Central exclusive production (CEP) of high-mass systems,
suggested almost twenty years ago in \cite{originally},
has recently become one of the hot topics in hadronic diffraction.
The main interest here is the possibility to detect the Higgs boson at the LHC
in a very clean environment. If the forward detectors, such as those
developed by the FP420 collaboration \cite{FP420}, can measure the energy of the scattered
protons, the mass of the central system
can be reconstructed with high accuracy and the Higgs signal
can be seen.

In the last decade several attempts have been made to predict the Higgs
CEP, with results differing by orders of magnitude \cite{various}.
Although there is now a broad consensus on the general structure of the CEP amplitude,
debates continue on how to implement the various ingredients.
CDF has recently published their data on central production of dijets
\cite{CDF}, which can be used to contrain theoretical models of CEP and, hopefully,
to make the estimates of the Higgs CEP cross section at the LHC more reliable.

Recently, we calculated the cross section of dijet CEP, see details in \cite{paper}.
We followed the standard line of calculations, trying, however, to keep an eye on
various assumptions and uncertainties that arise along the way.
We came to the conclusion that the uncertainties are much larger than
usually stated, and that the claim of the domination of the perturbative contribution
is not justified.
Besides, we argued that the link between the dijet and Higgs CEP
might be not as direct as it is often assumed.

\section{The standard scheme for hard dijet CEP}

The standard scheme for the calculation of CEP of system $X$ consists
in the following steps: (1) calculate the Born-level amplitude $qq\to q+X+q$,
(2) put quarks into protons via the proton form factors,
(3) take into account double-log enhanced corrections via the Sudakov form factor;
(4) suppress inelastic rescattering by introducing a gap survival probability.
For dijet production, one has in addition to convert produced gluons into final jets.

One begins with the lowest-order pQCD calculation of the $qq \to q+gg+q$ amplitude,
where the two-gluon system is a color singlet and there is no overall color exchange in the $t$-channel.
The kinematical conventions are the following (see Fig.~1):
the transverse momentum transfers to the two protons, $\bk_1$
and $\bk_3$, are small, while the relative momentum of the two gluons, $\bk_2$, is large.
The gluons are produced at mid-rapidity, so that the lightcone components
of their momenta are much smaller than the corresponding quark components.

\begin{figure}
  \includegraphics[height=.15\textheight]{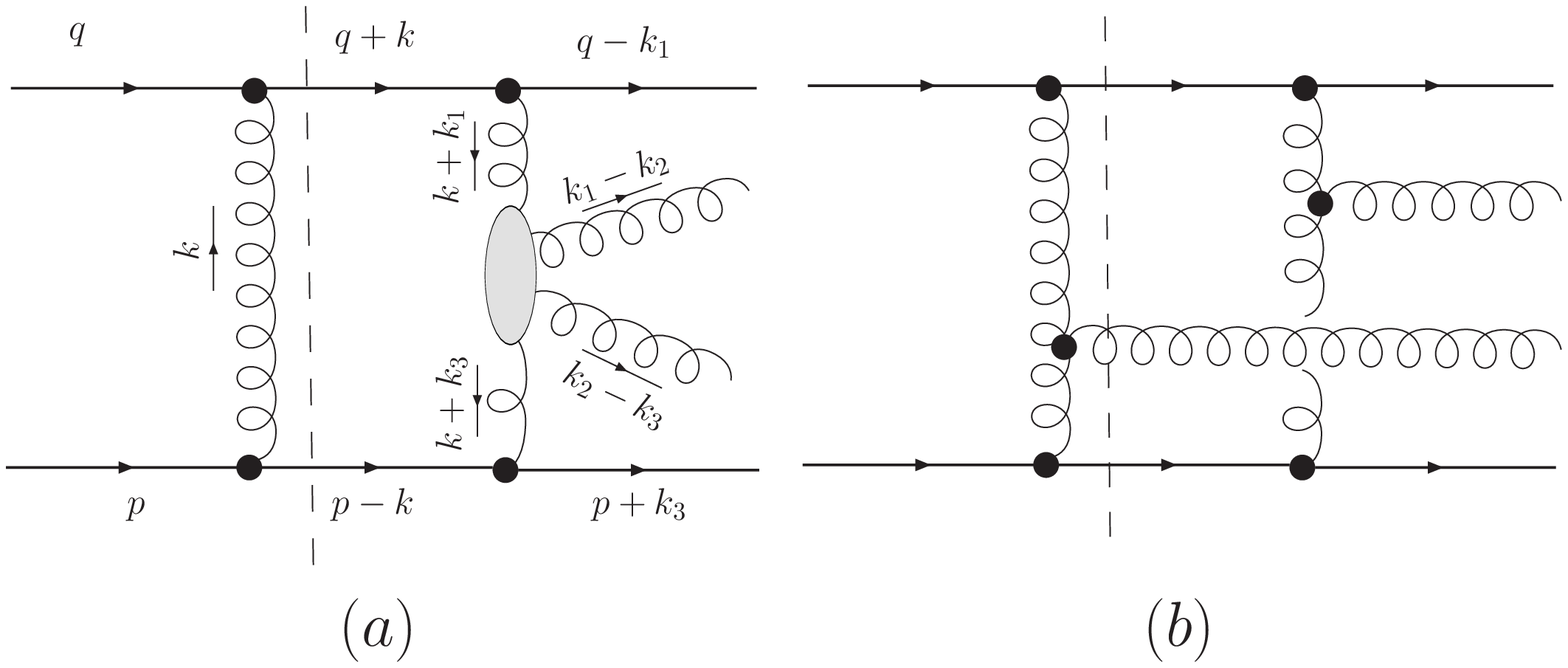}
  \caption{Diagrams contributing to the CEP of hard gluons}
\end{figure}

There are many diagrams contributing to the amplitude. However, if we focus on the imaginary part,
which is expected to be dominant, and take into account kinematics, we observe massive cancellations
among contributions. What remains is the standard diagrams with the two gluons emitted from
the same $t$-channel gluon, such as in Fig.~1a. The cross section can then be written as
\bea
d\sigma &=& {\alpha_s^6 \over 8 \pi^6 } \,{N^2-1 \over N^2} {1 \over (\bk_2^2)^2}
\,{d\beta_1 \over\beta_1}{d\beta_2 \over\beta_2}\, d^2\bk_1\,d^2\bk_2\,d^2\bk_3\nonumber\\
&&\hspace{-15mm}\times \int {d^2\bk\ d^2\bk^\prime \over \bk^2 (\bk+\bk_1)^2 (\bk+\bk_3)^2\
\bk^{\prime 2} (\bk^\prime+\bk_1)^2 (\bk^\prime+\bk_3)^2}
(C_0 |\tilde{M_0}|^2 + C_2 |\tilde{M_2}|^2)\,,\label{born}
\eea
where $\beta_i$ are the fractions of lightcone momentum carried by the two produced gluons,
$C_0$, $C_2$ are some products of transverse momenta, while $\tilde{M_0}$ and
$\tilde{M_2}$ are $gg \to gg$ amplitudes with total helicity 0 and 2 (see details in \cite{paper}).

Two gluon production in the collision of two protons
can be written schematically as
$d\sigma_{pp} = d\sigma_{qq} \otimes \prod_i \left[\sqrt{T_i}\cdot N_c \Phi_i\right]$.
Here $\Phi_i$ are the proton form factors, $T_i$ are the Sudakov form factors,
and the symbol $\otimes$
indicates that the corresponding factors are introduced inside the gluon loop integrals.
The proton form factors are usually rewritten via unintegrated gluon distribution functions.
We used the parametrizations of \cite{IN2002} adapted
for non-forward kinematics. As for the Sudakov form factor, we used the standard expression based
on the Monte Carlo approach to parton branching:
$$
T = \exp(-S)\,,\quad
S(\mu^2,\ell^2)=\int_{\ell^2}^{\mu^2} {d\bq^2 \over \bq^2} {\alpha_s(\bq^2) \over 2\pi}
\int_0^{1-\Delta} dz \left[zP_{gg}+N_f P_{gq}\right]\,.\label{sudasplitting}
$$
Here the lower scale $\ell^2$ is the virtuality of the colliding gluons,
the upper scale $\mu^2 \sim \bk_2^2$; $\Delta$ is a kinematic cut-off,
$P_{gg}$ and $P_{gq}$ are the non-regularized splitting functions.

Finally, we used the gap survival probability of 15\%, though a more detailed
analysis is definitely needed, and took into account the splash-out effects
via two prescriptions: $E_{T}^{j} = 0.8 k_T^g$ or
$E_T^j = k_T^g(1-\alpha_s/2)- 1\, \mathrm{GeV}$.

\section{Uncertainties of the calculations}

We followed the above steps to get numerical results for the dijet production cross section
within the kinematical cuts used by the CDF collaboration. We found that with appropriate choices
of free parameters we could nicely describe the CDF data. The corresponding curve is
represented in Fig.~2 by the solid line.

\begin{figure}
  \includegraphics[height=.3\textheight]{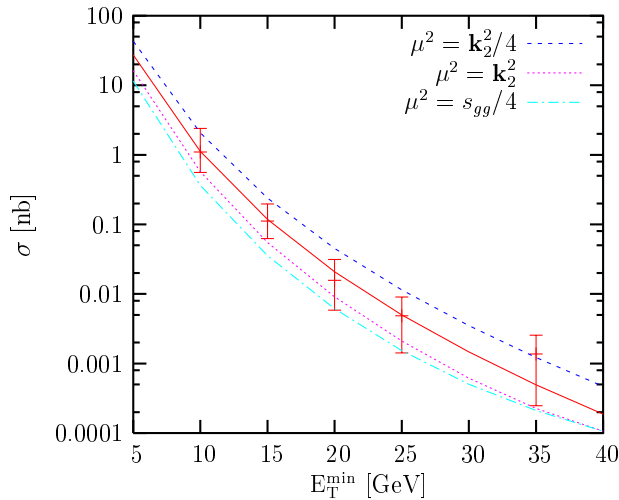}
  \caption{Sensitivity of the differential cross section to the choice of the upper scale
  $\mu^2$ in the Sudakov form factor}
\end{figure}

However, we also found that there are several issues that make the
numerical calculations very uncertain. For example, we found that
different choices of the unintegrated gluon distribution,
which did an equally good job in describing
the HERA data on $F_{2p}(x,Q^2)$ in very broad $x$ and $Q^2$ ranges,
lead to dijet CEP differing by a factor of 2--3.
This result is not surprising, since
the CEP cross section is proportional to the fourth power of the unintegrated gluon distribution function
and the CEP is dominated by the non-perturbative contribution.
Indeed, we checked that only $\sim 1/3$ of the cross section comes from kinematical configurations in which
all gluons inside the loop have virtualities larger than 1 GeV$^2$.

We also question the accuracy with which the Sudakov form factor is known.
First, we observed that for the CDF kinematics the constant terms in the Sudakov integral
are equally important as double-log plus single-log terms.
Variation of the constant terms, which are not known for the virtual corrections,
strongly affects the value of the Sudakov form factor.
In addition, reasonable variations of the scales of the Sudakov integral lead
to a significant
spread of the resulting numerical calculations,
which is illustrated by Fig.~2.

We estimated the cumulative uncertainty due to these and similar
sources to be a factor of 20 up or down,
which is much larger than usually claimed.
Therefore, the CDF data are indeed very important in constraining the models.

In addition to the uncertainties due to parametrizations, we identified diagrams,
which are usually neglected but which are not under theoretical control.
Consider, for example, the diagram in Fig.~1b. At the Born level, it is suppressed
with respect to the leading diagram by a power of $\bk_2^2$. However, these
diagrams get very large corrections due to logarithmically enhanced higher-order diagrams.
After these corrections are taken into account, the two diagrams shown in Fig.~1 compare
to each other roughly as
\be
{1\over \bk_2^2} \int {d^2\bk \over (\bk^2)^2} \exp[-S(\bk_2^2,\bk^2)]\quad
\mathrm{vs.}\quad
{1\over (\bk_2^2)^2} \int {d^2\bk \over \bk^2} \exp[-S_{new}(\bk_2^2,\bk^2)]\,.
\ee
Here $S_{new}$ is an effective way of representing enhanced virtual corrections to the
second diagram. We do not have an expression for $S_{new}$, but we can argue that it is smaller
(i.e. less suppressive) than the usual Sudakov integral $S$.
Note also that the second integral extends to much harder $\bk$.
Therefore, it might happen that after the corrections are taken into account,
the two diagrams can be comparable to each other.
Since the diagram shown in Fig.~1b is specific for $gg$ production and is absent in Higgs CEP,
one must bring it under control before claiming the similarity between the dijet and Higgs CEP.
In addition to this $gg$-specific correction, we have also identified
a class of diagrams that might be potentially important
for any final state in CEP, see details in \cite{paper}.

In conclusion, we argue that the calculations of the dijet CEP
are much more uncertain than usually claimed.
We find large uncertainties due to
the dominance of the soft dynamics, freedom in choosing parametrizations for the
unintegrated gluon distribution functions, strong sensitivity to the
details of the Sudakov form factor and, potentially, additional diagrams.
Even with the CDF data taken into account, estimates
of the Higgs CEP at the LHC remain rather uncertain.


\begin{theacknowledgments}
We thank M.Ryskin, V.Khoze, and A.Papa for useful discussions at various stages of this work.
The work of I.P.I. was supported by the Belgian Fund F.R.S.-FNRS via the contract of
Charg\'e de recherches.
\end{theacknowledgments}



\bibliographystyle{aipprocl} 




\end{document}